\documentclass[aps,twocolumn,nofootinbib,superscriptaddress]{revtex4-1}
\usepackage{amsfonts,ulem,bbold}
\usepackage{amssymb,amsmath}
\usepackage{color}
\usepackage{epsfig}

\newcommand{\Comment}[1]{{}}
\definecolor{MyDarkBlue}{rgb}{0.15,0.15,0.45}
\usepackage[linktocpage=true]{hyperref}
\hypersetup{
colorlinks=true,
citecolor=MyDarkBlue,
linkcolor=MyDarkBlue,
urlcolor=MyDarkBlue,
}

\newcommand{\be}{\begin{equation}}
\newcommand{\ee}{\end{equation}}
\newcommand{\bea}{\begin{eqnarray}}
\newcommand{\eea}{\end{eqnarray}}
 
\newcommand{\e}{{ \rm{e}}}

\begin{document}

\title{Black Hole Mechanics for Massive Gravitons}

\author{Rachel A. Rosen}
\affiliation{Department of Physics, Columbia University,\\ New York, NY 10027, USA}

\begin{abstract} 

It has been argued that black hole solutions become unavoidably time-dependent when the graviton has a mass.  In this work we show that, if the apparent horizon of the black hole is a null surface with respect to a fiducial Minkowski reference metric, then the location of the horizon is necessarily time-{\it independent}, despite the dynamical metric possessing no time-like Killing vector.  This result is non-perturbative and model-independent.  We derive a second law of black hole mechanics for these black holes and determine their surface gravity.  An additional assumption establishes a zeroth and a first law of black hole mechanics.  We apply these results to the specific model of dRGT ghost-free massive gravity and show that consistent solutions exist which obey the required assumptions.  We determine the time-dependent scalar curvature at the horizon of these black holes.

\end{abstract}

\maketitle

\section{Introduction and Summary}
\label{intro}
 \vspace{-.5cm}
It remains an open question: what would happen to a black hole if the graviton were to have a mass?  Static black hole solutions in massive gravity appear to unavoidably suffer from either infinite strong coupling at all scales or coordinate-invariant curvature singularities at the horizon.  (See, e.g., \cite{Deffayet:2011rh,Mirbabayi:2013sva,Rosen:2017dvn} for specific arguments.)  In \cite{Mirbabayi:2013sva,Rosen:2017dvn} it was argued that these pathologies can be avoided by adopting a time-dependent, spherically symmetric ansatz.  In particular, perturbative time-dependent solutions were found in \cite{Rosen:2017dvn} which are regular at the horizon, are  potentially consistent with the expected Yukawa asymptotics at large distances and weak coupling, and possess a massless limit that smoothly recovers the Schwarzschild black holes of General Relativity.  However, finding exact analytic time-dependent black hole solutions has proved challenging.

In this work, we derive non-perturbative results for these time-dependent black holes by focusing on the horizon.  We are primarily interested in the laws of black hole mechanics \cite{Bardeen:1973gs} in the presence of a graviton mass.  In general, black hole mechanics are non-trivial for time dependent metrics.  We refer to \cite{Hayward:1994bu,Hayward:1997jp,Ashtekar:2004cn,Nielsen:2005af,Nielsen:2007ac} for resources.  

We first consider general theories of a massive spin-2 particle.  Generic theories of massive gravitons require a fiducial reference metric $f_{\mu\nu}$ in addition to the dynamical metric $g_{\mu\nu}$ in order to construct non-derivative potential terms.  In this paper we show that, if the apparent horizon of the black hole is a null surface with respect to a fiducial Minkowski reference metric, then the location of the apparent horizon is time-{\it independent} in the coordinates of the dynamical metric.  We derive several additional consequences of this initial assumption: the apparent horizon is then also a null surface in terms of the dynamical metric; the surface gravity can be computed in terms of the inaffinity of the Kodama vector at the horizon; and the area of the apparent horizon is never decreasing, consistent with a second law of black hole mechanics.  An additional assumption is required so that the surface gravity is constant in time and thus these black holes possess a zeroth law as well.  A first law follows naturally.

We then consider the implication of these results for dRGT ghost-free massive gravity \cite{deRham:2010kj}.  We solve the dRGT equations of motion in the vicinity of the horizon.  We find solutions consistent both with the assumption of a null apparent horizon and that have a time-independent surface gravity.  These solutions are nevertheless truly time-dependent in that they possess no time-like killing vector.  In particular, the scalar curvature is not constant along the horizon.  We compute the time-dependent scalar curvature at the horizon and discuss some implications.

\section{Fixed Horizons}
\label{horizons}
 \vspace{-.5cm}
To construct a theory of a massive spin-2 particle, one adds a non-derivative potential term to the Einstein-Hilbert Lagrangian.  In order to contract the indices of the dynamical metric $g_{\mu\nu}$ in the potential in a non-trivial way, one needs to introduce an additional fiducial ``reference" metric $f_{\mu\nu}$.  In order for the theory to be Lorentz-invariant, this reference metric is taken to be Minkowski $f_{\mu\nu} = \eta_{\mu\nu}$ (or a coordinate transformation thereof).  The most general potential can be written as the trace of various powers of the matrix $g^{-1}f = g^{\mu\lambda}f_{\lambda\nu}$.  We can express this as:
\be
\label{U}
U(g,f) =\sum_{n \in \mathbb Q} c_n\, {\rm tr} \left[ (g^{-1}f)^n \right] \, .
\ee
Here, the $c_n$ are arbitrary constant coefficients and we only assume that $n$ is a rational number, i.e., we include fractional powers and inverse powers of $g^{-1}f$.  A generic massive gravity Lagrangian can thus be written as
\be
\label{Lgen}
{\cal L} = \frac{M_{Pl}^2}{2}\sqrt{-g}\left[R(g)-m^2\, U(g,f) \right] \, ,
\ee
where $R$ is the usual Ricci scalar and $m$ is the graviton mass.

The reference metric is Minkowski.  Thus there exists a set of coordinates which we denote by $(\tau,\rho,\theta,\phi)$ in which the line element for $f_{\mu\nu}$ takes the form  
\be
\label{f}
ds^2_f = -d\tau^2+d\rho^2+\rho^2 d\Omega^2 \, .
\ee
For the dynamical metric $g_{\mu\nu}$, we consider time-dependent, spherically symmetric vacuum solutions.  There exists a set of coordinates which we denote by $(t,r, \theta, \phi)$ in which a generic spherically symmetric metric can be written as 
\be
ds^2_g =\! -\e^{2 \Phi(t,r)} \left(1- \tfrac{2 G M(t,r)}{r} \!\right) dt^2+\frac{dr^2}{1- \frac{2 G M(t,r)}{r}} +r^2 d\Omega^2 \, .
\ee
The functions $\Phi(t,r)$ and $M(t,r)$ are arbitrary and will be fixed by the equations of motion.  We have chosen this definition of the radial coordinate $r$ so that spheres have area given by $A=4\pi r^2$.   In general these coordinates $(t,r)$ will not coincide with the $(\tau,\rho)$ coordinates defined above.  However, at large distances from a source, we expect the dynamical metric to be approximately flat.   In order for our Lagrangian to recover the Fierz-Pauli form at weak coupling, we expect $t \rightarrow \tau$ and $r \rightarrow \rho$ at large $r$.

The time coordinate $t$ for the dynamical metric is ill-defined when $2 G M(t,r) = r $.  It is useful to adopt a null, ingoing Eddington-Finkelstein type coordinate $v$ instead.  The dynamical metric becomes
\be
\label{ansatz}
ds^2_g =\! -F(v,r)^2 {\scriptstyle{ \left(1- \tfrac{2 G {\cal M}(v,r)}{r} \!\right)}} dv^2+2F(v,r)dv dr +r^2 d\Omega^2 \, .
\ee
where ${\cal M}\!\left[v(t,r),r\right] = M(t,r)$ and $\Phi(t,r)$ has been absorbed into a new function $F(v,r)$.  In this work we refer to the usual three-dimensional trapping horizon (i.e., the union of the 2D boundaries of the trapped regions) as the apparent horizon.  In the coordinates given above \eqref{ansatz}, the location of the apparent horizon is determined by the implicit expression
\be
\label{rH}
2 G {\cal M}(v,r_H) = r_H ~~ \Rightarrow ~~ r_H(v) \, .
\ee
For a generic function ${\cal M}(v,r)$, the location of the horizon will depend on the lightcone coordinate $v$.  In general, the apparent horizon will not be a null surface.  The norm of the vector normal to the horizon $n^\mu$ is given by
\be
\label{null}
n^\mu n_\mu = -\frac{2\,  \dot{r}_H(v)}{F(v,r_H(v))} \, ,
\ee
where dot denotes derivative with respect to $v$.  Thus only when the location of the apparent horizon is independent of $v$ will the horizon be a null surface with respect to the dynamical metric.  What's more, null surfaces of the two metrics $f_{\mu\nu}$ and $g_{\mu\nu}$ do not generically coincide.  Thus, even if the the apparent horizon is a null surface with regards to $g_{\mu\nu}$, it will not necessarily be so with respect to $f_{\mu\nu}$.

The reference metric coordinates $(\tau,\rho)$ are related to the Eddington-Finkelstein coordinates $(v,r)$ via a yet-undetermined coordinate transformation:
\be
\label{transf}
\tau = \tau(v,r)\, , ~~~~ \rho = \rho(v,r) \, .
\ee
The four independent equations of motion that follow from the Lagrangian \eqref{Lgen} fix the four unknown functions: ${\cal M}(v,r)$, $F(v,r)$, $\tau(v,r)$ and $\rho(v,r)$.

Let us consider explicitly the equations of motion using the ansatz \eqref{ansatz} for the dynamical metric.  The variation of the potential can be found using the relation
\be
\delta\, {\rm tr}\! \left[ (g^{-1}f)^n \right] 
= n\,  {\rm tr}\! \left[ g\, (g^{-1}f)^n \delta g^{-1} \right] \, .
\ee
Accordingly, the vacuum equations of motion are 
\be
\label{GU}
R^\mu_{~\nu} -\tfrac{1}{2} \delta^\mu_{~\nu} R - m^2 U^\mu_{~\nu}  =0 \, , 
\ee
where
\be
U^\mu_{~\nu} \equiv  \sum_{n \in \mathbb Q} c_n \left( n \left[(g^{-1}f)^n\right]^\mu_{~\nu} - \tfrac{1}{2}\delta^\mu_{~\nu}  {\rm tr} \! \left[ (g^{-1}f)^n \right]   \right) \, .
\ee

We are particularly interested in the off-diagonal $(1,0)$ component of the equations of motion which is given by
\be
\label{eom}
\frac{2\,G}{r^2}\, \dot{{\cal M}}(v,r) = m^2  \sum_{n \in \mathbb Q} c_n  n \left[(g^{-1}f)^n\right]^1_{~0} \, .
\ee
In order to evaluate the right hand side of the above expression we note that, for a $2\times 2$ matrix, the off-diagonal components of the matrix raised to some power $n$ (including fractional or inverse powers) are always proportional to the same off-diagonal component of the matrix itself.  I.e., if $\mathbb{X}$ is a $2\times 2$ invertible matrix then
\be
\left(\mathbb{X}^n\right)_{12} = \mathbb{X}_{12} \times  (\ldots) \, ,~~~~ \left(\mathbb{X}^n\right)_{21} = \mathbb{X}_{21}  \times  (\ldots) \, .
\ee
Since the matrix $g^{-1}f$ is diagonal in its angular components, it can be expressed as the direct product of two $2\times 2$ matrices and the above argument can be applied.  In other words, 
\be
\label{prop}
\left[(g^{-1}f)^n\right]^1_{~0} = (g^{-1}f)^1_{~0}  \times (\ldots) \, .
\ee
For the metrics defined above in \eqref{f} and \eqref{ansatz}, this is component is determined to be
\begin{align}
\label{gf}
&(g^{-1}f)^1_{~0} = \frac{\dot{\rho}(v,r)^2-\dot{\tau}(v,r)^2}{F(v,r)}  \\
&+\left(1-\tfrac{2 G {\cal M}(v,r)}{r}\right) \left(\dot{\rho}(v,r)\rho'(v,r)-\dot{\tau}(v,r)\tau'(v,r)\right) \, , \nonumber
\end{align}
where dots denote derivatives with respect to $v$ and primes denote derivatives with respect to $r$.

Let us now make an assumption that a null apparent horizon is a desirable feature for our black hole to have, in particular because it means that the apparent horizon might be identified with the event horizon of the black hole once the full structure of the spacetime is known.  Let us also use the external structure of the reference metric to define the ``nullness" of the horizon.   In other words, let us consider the consequences if we assume that
\be
\tau(v,r_H(v)) = \pm \rho(v,r_H(v))+ const\, .
\ee
An immediate result is that the rhs of \eqref{gf} will vanish when evaluated at the horizon
\be
\left. (g^{-1}f)^1_{~0} \right|_{r=r_H} = 0 \, .
\ee
From the property \eqref{prop} and the equations of motion \eqref{eom} it follows that
\be
\label{Mdot}
\dot{{\cal M}}(v,r_H) = 0 \, .
\ee
The location of the horizon is thus time-independent
\be
r_H(v) = const \, .
\ee
We note that this is a coordinate invariant statement in the following sense: if we fix the radial coordinate so that spheres have the canonical area given by $A=4\pi r^2$, then no coordinate transformation $v$ will alter this statement of time-independence.

Furthermore, referring to equation \eqref{null}, we see that if $\dot{r}_H(v) = 0$ then the apparent horizon is a null surface in terms of the dynamical metric as well.  In other words, for the generic massive gravity theory, assuming that the apparent horizon was null with respect to the reference metric also implied its nullness with respect to the dynamical metric.  In the next section we explore additional implications for black hole mechanics.

\section{Mechanics}
\label{mechanics}
 \vspace{-.5cm}
We now consider the consequences of the graviton mass on black hole mechanics.   We start with the second law.  We construct an outward radial null vector
\be
k^\mu = N \left(\frac{2}{F(v,r)},1-\frac{2 G{\cal M}(v,r)}{r},0,0 \right) \, ,
\ee
up to some overall normalization $N$.  Let us introduce a source $T_{\mu\nu}$ to the equations of motion \eqref{GU} and contract all indices with $k^\mu$:
\be
k_\mu G^\mu_{~\nu}k^\nu - m^2 k_\mu U^\mu_{~\nu} k^\nu =8 \pi G\, T_{\mu\nu}k^\mu k^\nu \, .
\ee
Evaluating this expression at the horizon gives
 \be
 \label{source}
\left.  \frac{4 N^2}{F(v,r)} \left(G^1_{~0}-m^2 U^1_{~0} \right) \right|_{r=r_H} = \left. 8 \pi G\, T_{\mu\nu}k^\mu k^\nu \right|_{r=r_H}\, .
 \ee
 As argued in the previous section, $U^1_{~0}$ vanishes at the horizon whenever the apparent horizon is a null surface in terms of $f_{\mu\nu}$.  If this condition holds for the sourced equations of motion, then the term arising from the graviton mass drops out of the above expression \eqref{source}.  We are left with the usual expression for General Relativity (see, e.g., \cite{Nielsen:2005af})
 \be
\frac{4 N^2}{F(v,r_H(v))}  \frac{2 G \dot{{\cal M}}(v,r_H(v)) }{r_H(v)^2}= \left. 8 \pi G\, T_{\mu\nu}k^\mu k^\nu \right|_{r=r_H}\, .
 \ee
As we have defined $v$ in equation \eqref{ansatz} to be an {\it ingoing} null coordinate, the function $F(v,r)$ is positive.  It follows that
 \be
 \dot{r}_H(v)  \geq 0 \, ,
\ee
 as long as
\be
T_{\mu\nu}k^\mu k^\nu \geq 0 \, .
\ee
In other words, the area of the apparent horizon will not decrease as a function of time (i.e., $v$) as long as the null energy condition is satisfied.

To determine the surface gravity of these black holes, we use the Kodama vector \cite{Kodama:1979vn} in place of the time translation Killing vector used for static and stationary spacetimes:
 \be
K^\mu = \frac{1}{F(v,r)} \left(1,0,0,0 \right)\, .
 \ee
The norm of the Kodama vector is given by
\be
K^\mu  K_\mu = - \left(1-\frac{2 G {\cal M}(v,r)}{r} \right) \, .
\ee
Like the time translation Killing vector, $K^\mu$ is normalized so that $K^\mu  K_\mu \rightarrow -1$ as $r \rightarrow \infty$, assuming our dynamical metric has flat asymptotics.  Furthermore, the Kodama vector is null along the black hole horizon.

If we adopt the condition of a fixed horizon \eqref{Mdot} (and only by adopting this condition), we find that $K^\mu$ obeys the geodesic equation in a non-affine parametrization at the horizon:
\be
 \left. K^\mu \nabla_\mu K^\nu \right|_{r=r_H}  =\left.   \kappa\, K^\nu  \right|_{r=r_H} \, ,
 \ee
 where 
 \be
 \label{kappa}
 \kappa \equiv \frac{1-2 G {\cal M}'(v,r_H)}{2 \, r_H} \, .
 \ee
Again, prime denotes derivative with respect to $r$.  We define the surface gravity $\kappa$ via the inaffinity of $K^\mu$.

The zeroth law of black hole mechanics states that the surface gravity $\kappa$ is constant everywhere along the horizon, i.e., as a function of $v$.  Here we see that this statement holds only when ${\cal M}'(v,r_H) = 0$.  If this condition is assumed, by differentiating ${\cal M}(v,r_H(v))$ with respect to $v$, it is straightforward to show that a first law of black hole mechanics is also obeyed
\be
\label{first}
d{\cal M}= \frac{\kappa}{8\pi}d{\cal A} \, ,
\ee
where ${\cal A}=4\pi r_H^2$.  The function ${\cal M}(v,r)$ the Misner-Sharp mass function \cite{Misner:1964je} which can be interpreted as the mass inside radius $r$ as a function of $v$.  It is equivalent to the charge associated with the conserved current arising from the Kodama vector; it reduces to the Bondi-Sachs energy \cite{Bondi:1962px,Sachs:1962wk} at null infinity and  the Arnowitt-Deser-Misner energy \cite{Arnowitt:1962hi} at spatial infinity (see \cite{Hayward:1994bu}).

For generic massive gravity, the function ${\cal M}'(v,r)$ is fixed by the $(0,0)$ component of the equations of motion
 \be
-\frac{2 G {\cal M}'(v,r)}{r^2} =  m^2  \sum_{n \in \mathbb Q} c_n \left( n \left[(g^{-1}f)^n\right]^0_{~0} - \tfrac{1}{2} {\rm tr} \! \left[ (g^{-1}f)^n \right] \right) \!, 
 \ee
and is not necessarily zero at the horizon.  In the next section we will see that, in the specific case of ghost-free massive gravity, solutions exist with ${\cal M}'(v,r_H) = 0$.

 \section{Ghost-Free Massive Gravity}
\label{dRGT}
 \vspace{-.5cm}
We now consider the implications of the above discussion for  dRGT ghost-free massive gravity \cite{deRham:2010kj}.  The relevant Lagrangian is given by
\be
\label{L}
{\cal L} = \frac{M_{Pl}^2}{2}\sqrt{-g}\left[R-2 m^2\sum_{n=0}^4 \beta_n S_n(\sqrt{g^{-1} f}) \right] \, .
\ee
In the potential term, the $S_n$ are the $n$-th elementary symmetric polynomials of the eigenvalues of the matrix square root of $g^{\mu\lambda}f_{\lambda\nu}$.  They are given by
\bea
\label{potential}
\begin{array}{l}
S_0 (\mathbb{X})= 1  \, ,  \vspace{.1cm} \\
S_1(\mathbb{X})= [\mathbb{X}]  \, ,  \vspace{.1cm} \\
S_2(\mathbb{X})= \tfrac{1}{2}([\mathbb{X}]^2-[\mathbb{X}^2]) \, ,  \vspace{.1cm} \\
S_3(\mathbb{X})= \tfrac{1}{6}([\mathbb{X}]^3-3[\mathbb{X}][\mathbb{X}^2]+2[\mathbb{X}^3]) \, ,  \vspace{.1cm} \\
S_4(\mathbb{X})=\tfrac{1}{24}([\mathbb{X}]^4-6[\mathbb{X}]^2[\mathbb{X}^2]+3[\mathbb{X}^2]^2   
+8[\mathbb{X}][\mathbb{X}^3]-6[\mathbb{X}^4])\, .
\end{array}
\eea
Here (and only here) the square brackets denote the trace of the enclosed matrix.   This form of the potential guarantees that the classical theory propagates only the correct five degrees of freedom of the massive graviton \cite{Hassan:2011hr} and no additional Boulware-Deser ghost \cite{Boulware:1973my}.  Though somewhat unwieldy in metric notation, the polynomials have a more elegant rewriting as wedge products of vielbeins \cite{Hinterbichler:2012cn}.

The $\beta_n$ are free coefficients.  If we expand the dynamical metric around flat spacetime $g_{\mu\nu} = \eta_{\mu\nu} + 2 h_{\mu\nu}/M_{Pl}$ then the requirement of no tadpoles gives the following condition on the $\beta_n$:
\be
\label{cc}
\beta_0 + 3 \beta_1 + 3 \beta_2 + \beta_3 = 0 \, .
\ee
In other words, this condition sets to zero the cosmological constant term coming from the potential.  Assuming this condition, the correct normalization of the mass $m^2$ means that
\be
\label{mass}
\beta_1 +2\beta_2 +\beta_3 =1 \, .
\ee
This condition guarantees that around flat spacetime, at lowest order in the fields, the Lagrangian \eqref{L} reduces to the linear Fierz-Pauli Lagrangian for the free massive graviton, with mass term 
\be
\label{FP}
 \frac{m^2}{2}(h_{\mu\nu}h^{\mu\nu}-h^\mu_{\, \mu}h^\nu_{\, \nu}) \, .
\ee

We adopt the time-dependent spherically symmetric ansatz given above \eqref{ansatz} and solve the vacuum equations of motion that follow from \eqref{L} in the vicinity of the horizon.   Here, for clarity, we present only the relevant boundary values of the solutions.  We find consistent solutions that obey the condition for a null horizon:
\be
\label{cond}
V(v) \equiv \tau (v,r_H) = \rho(v,r_H)  \, . 
\ee
The function $V(v)$ has the interpretation of the ingoing lightcone coordinate for the Minkowski reference metric on the black hole horizon:
\be
2\, V(v) = \tau (v,r_H) + \rho(v,r_H)   \,.
\ee
(Note that $V \geq 0$ as $\rho$ is a radial coordinate and thus $V(v)=\rho(v,r_H)$ is always positive semidefinite.)   With the condition \eqref{cond}, it follows that we can set the location of the horizon to be independent of $v$:
\be
 {\cal M}(v,r_H) =\frac{r_H}{2\,G} \, .
\ee
We also set the surface gravity to be constant everywhere on the horizon:
\be
 {\cal M}'(v,r_H) = 0 \, .
\ee

After setting these boundary conditions, the remaining equations of motion near the horizon give the following relation between $v$ and $V$:
\be
\label{Vsol}
2\, V(v) = 2 \, r_H\,  \e^{v/2 r_H} \, ,
\ee
We can compare this coordinate transformation to the transformation between lightcone Rindler $v_{\rm R}$ and lightcone Minkowski $V_{\rm M}$ coordinates:
\be
V_{\rm M} = \tfrac{1}{a} \e^{a v_{\rm R}} \, ,
\ee
where $a$ is the Rindler acceleration.  Thus, by analogy, we can identify the surface gravity $\kappa = 1/2r_H$ as we would expect from the previous arguments.  (The factor of two on the lhs of \eqref{Vsol} is a matter of convention and follows from our definition \eqref{cond}.) 

The function $F(v,r)$ evaluated at the horizon is also determined.  It simplifies when expressed in terms of $V$.  We find:
\be
F[v(V),r_H] = \frac{V\,f(V)}{\int\! dV\,f(V)} \, , 
\ee
where
\be
f(V) \equiv \left(\frac{\beta_1 r_H^2+2  \beta_2 r_H  V+\beta_3 V^2}{\beta_0 r_H^2 +2 \beta_1 r_H  V+\beta_2V^2}  \right)^{\!2}  \, .
\ee
With these boundary solutions, the equations of motion can be solved perturbatively, order by order in $r-r_H$ to find ${\cal M}(v,r)$, $F(v,r)$, $\tau(v,r)$ and $\rho(v,r)$ away from the horizon.

To summarize, we find that dRGT massive gravity possesses time-dependent, spherically symmetric vacuum solutions where (i) the apparent horizon is a null surface and thus the location of the horizon is time-independent and (ii) the surface gravity is everywhere constant on the apparent horizon.  However, the metric still possesses no time-like Killing vector.  Other physical quantities evaluated on the horizon will generically depend on $v$ (or, equivalently, $V$).

In particular, we can compute the scalar curvature of the dynamical metric on the horizon as a function of the ingoing lightcone coordinate $V$ on our solution.  We find
\begin{align}
&R(V) = - \frac{2 m^2}{r_H}(\beta_0 r_H^3  + 3 \beta_1 r_H^2 V  + 3 \beta_2 r_H V^2  + \beta_3 V^3 ) \nonumber \\ 
&\times \frac{(\beta_1^2 - \beta_0 \beta_2)r_H^2  +  (\beta_1 \beta_2 - \beta_0 \beta_3)  r_H V +  (\beta_2^2 - \beta_1 \beta_3)V^2}
{( \beta_1 r_H^2+2 \beta_2 r_H  V+\beta_3V^2 )^2} \, . \nonumber \\ 
\end{align}
This function exhibits the following curious properties.  When $v=0$ we have $V=r_H$ and $R =0$ for all values of the $\beta_n$.  We can set $R(V) = 0$ for all $V$ if we choose the $\beta_n$ such that
\be
\beta_0  = \frac{\beta_1^2}{\beta_2}  \, , ~~~~  \beta_3  =\frac{\beta_2^2}{\beta_1} \, .
\ee
However, these conditions cannot be satisfied if both \eqref{cc} and \eqref{mass} are also satisfied.  Thus the curvature at the horizon is generically nonzero away from $v=0$.

\section{Discussion}
\label{discuss}
\vspace{-.5cm}
Having a null apparent horizon is a potentially attractive feature for a time-dependent black hole.  It means that the apparent horizon might coincide with the event horizon once the full structure of the spacetime is known.  In this work we have shown that for massive gravity, by assuming a null apparent horizon in terms of the fiducial reference metric, a number of additional desirable features are obtained.  In particular, we can guarantee (i) that the location of the horizon is time-independent (ii) that the apparent horizon is also null with respect to the dynamical metric (iii) the graviton mass term does not disrupt the usual second law of black hole mechanics and (iv) a notion of the surface gravity can be defined in terms of the inaffinity of the Kodama vector.

We have also shown that dRGT ghost-free massive gravity possesses solutions that realize this assumption of a null apparent horizon and that, moreover, have constant surface gravity and a first law of black hole mechanics given by \eqref{first}.  These solutions nevertheless possess a time-dependent scalar curvature at the horizon.  An immediate outstanding question is whether these solutions can be matched to the expected, static Yukawa asymptotics at large $r$.  Massive gravity has no Birkoff's theorem.  Thus it also remains to be seen if these black hole solutions can in fact be realized as, say, the endpoint of gravitational collapse.

Particularly in the age of LIGO, a better understanding of black hole solutions in massive gravity could potentially lead to improved constraints on the graviton mass.  Independent of observations, black holes in massive gravity are interesting in their own right: they possess horizons which aren't Killing horizons and the underlying theory has no diffeomorphism invariance.  Thus we cannot rely on the basic assumptions we frequently use to understand black holes in General Relativity (and related issues such as locality in GR).  A more thorough understanding of black hole thermodynamics and information for massive gravitons might ultimately teach us something interesting about black hole information in General Relativity.

\vskip.5cm

\bigskip
{\bf Acknowledgements}: 
I would like to thank Lam Hui, Austin Joyce, Janna Levin, Riccardo Penco and Robert Penna for discussions. This work was supported by DOE grant DE-SC0011941, NASA grant NNX16AB27G and Simons Foundation Award Number 555117.

\bibliographystyle{apsrev4-1}
\bibliography{blackholes}

\end{document}